# Static spherically symmetric solutions in general Gauss-Bonnet gravity

N. Mohammadipour[1]

Department of Physics Education, Farhangian University, P.O. Box 14665-889, Tehran, Iran.

**Abstract**

Considering an action in $F(G)$ modified gravity, the static spherically symmetric solutions are investigated. Introducing the Lagrangian multipliers $α$ we obtain the Lagrangian and equations of motion. we obtain two type solutions for these models. The first case leads to Schwarzschild-de Sitter (anti de Sitter) solution and the other one, results in a new metric. At last, the event horizon, the Hawking temperature and the generalized second law of thermodynamics in the framework of the modified Gauss-Bonnet gravity for this solution as a black hole are investigated.

# 1   Introduction

Recent accelerated expansion of our universe is one the most significant cosmological discoveries over the last decade [1, 2, 3, 4]. This acceleration is explained in terms of the so-called dark energy. Many candidates for the nature of dark energy have been proposed. The simplest suggestion for dark energy is cosmological constant. But it suffers from two kind of problems [5]: fine tuning and coincidence problem. A dynamical scalar field with quintessence or phantom behavior is another proposal for dark energy (for reviews see [6]). An alternative approach for the gravitational origin of dark energy is coming from modification of general relativity. Einstein s theory of gravity may not describe gravity at very high energy. The simplest alternative to general relativity is Brans-Dicke scalar-tensor theory [7]. But among the most popular modified gravities which may successfully describe the cosmic speed-up is $F(R)$ gravity [8, 9]. For this kind of modification, one assumes that the gravitational action may contain some additional terms which starts to grow with decreasing curvature and obtain a late time acceleration epoch. An alternative class of modified gravity models is the family of the string-inspired gravities by considering additional curvature invariant terms such as the Gauss-Bonnet (GB) term [10]. Nojiri et al [11] showed that a particular dark energy solution can be obtained from scalar-GaussBonnet cosmology. There is another version of the Gauss-Bonnet gravity namely the modified GB or $F(G)$ theory [12] that can also play the role of gravitational dark energy. Alternatively, one can consider the Lagrangian density as a general function of the Ricci scalar field and the Gauss-Bonnet invariant, $F(R,G)$ [13, 14].

---

[1] Corresponding email: naser.kurd@cfu.ac.ir

In this paper we investigate the static spherically symmetric solutions of $F(R,G)$ models. Here we use the method proposed in Ref.[15], for investigating static spherically symmetric solutions for a generic $F(R)$ model. The paper uses the signature (-,+,+,+) and repeated indices are to be summed. Units where $c = \hbar = k_B = G = 1$ are used unless explicitly stated otherwise. The layout of the paper is the following. In sec.(2), the action of $F(R,G)$ gravity is considered, and the field equations are derived, then the effect of modified gravity is considered as an effective stress-energy tensor. For the static spherically symmetric metric and using the method of Lagrangian multipliers to obtain the lagrangian and equations of motion is discussed in Sec.(3). In sec.(4), the static spherically symmetric vacuum solutions for a large class of these metrics are investigated and the exact solutions are obtained. Then we have obtained the event horizons, Hawking temperature and entropy of these solutions. Finally, sec.(5) is devoted to the discussion of our results.

## 2 Field equations and $F(R,G)$ modified gravity tensor

We start work by considering the action of $F(R,G)$- gravity in metric formalism as

$$S = \int d^4x \sqrt{-g}\left[\frac{1}{2}R + F(G)\right] + S_m, \qquad (1)$$

here $R$ is the scalar curvature, $F(G)$ is a function of $G$, where $G = R^2 - 4R_{\mu\nu}R^{\mu\nu} + R_{\mu\nu\alpha\beta}R^{\mu\nu\alpha\beta}$ is Gauss-Bonnet invariant, $R_{\mu\nu}$, $R_{\mu\nu\alpha\beta}$ are the Ricci and the Riemann tensors respectively, and $S_m$ is the matter action. The field equations for the Eq.(1) are [16]

$$G_{\mu\nu} = T_{\mu\nu} + 8[R_{\mu\rho\nu\sigma} + R_{\rho\nu}g_{\sigma\mu} - R_{\rho\sigma}g_{\nu\mu} - R_{\mu\nu}g_{\rho\sigma} + \qquad (2)$$

$$R_{\mu\sigma}g_{\nu\rho} + R/2(g_{\mu\nu}g_{\sigma\rho} - g_{\mu\sigma}g_{\nu\rho})]\nabla_\rho\nabla_\sigma F_{,G} - (GF_{,G} - F)g_{\mu\nu},$$

where $G_{\mu\nu}$ is the Einstein tensor, $T_{\mu\nu}$ is an energy momentum tensor of matter and $F_{,G} = \frac{dF(G)}{dG}$. For the special case of $F(G) \propto G$, we will recover the familiar Einstein field equations. By variation of action (1) with respect to the metric, we obtain

$$G_{\mu\nu} = R_{\mu\nu} - 1/2\, R g_{\mu\nu} = G_{\mu\nu}^{F(G)}, \qquad (3)$$

here, $G_{\mu\nu}^{F(G)}$ is an effective stress-energy tensor containing geometric terms as follows [16]

$$G_{F\mu\nu(G)} = -8[R_{\mu\rho\nu\sigma} + R_{\rho\nu}g_{\sigma\mu} - R_{\rho\sigma}g_{\nu\mu} - R_{\mu\nu}g_{\rho\sigma} + R_{\mu\sigma}g_{\nu\rho} + \qquad (4)$$

$$R/2(g_{\mu\nu}g_{\sigma\rho} - g_{\mu\sigma}g_{\nu\rho})]\nabla_\rho\nabla_\sigma F_{,G} - (GF_{,G} - F)g_{\mu\nu}.$$

In the next section, we would like to study the static spherically vacuum solution of the above model.

## 3 Static spherically symmetric vacuum solution

We take a generic form of metric for the static spherically symmetric solutions

$$ds^2 = -B(r)dt^2 + \frac{X(r)}{B(r)}dr^2 + r^2 d\theta^2 + r^2 \sin\theta^2 d\phi^2, \qquad (5)$$

where $B(r)$ and $X(r)$ are unknown functions of $r$.B The form of scalar curvature and Gauss-Bonnet term are as

$$R = -\frac{B''(r)}{X(r)} + \frac{B'(r)X'(r)}{2X^2(r)} - \frac{4B'(r)}{rX(r)} \qquad (6)$$
$$+ \frac{2B(r)X\prime(r)}{rX^2(r)} + \frac{2}{r^2} - \frac{2B(r)}{r^2X(r)}$$

$$G = -\frac{2}{r^2X(r)^3}[-2X(r)B(r)B''(r) - X(r)X'(r)B'(r)$$
$$+ 3X'(r)B(r)B'(r) + 2X'(r)B(r)B'(r)$$
$$+ 2X^2(r)B''(r) - 2X(r)B'^2(r)],$$

where $'$ and $''$ are first order and second order derivatives with respect to the radial coordinate $r$, respectively.

Introducing the Lagrangian multipliers $\alpha$ and making use of Eq.(6), we can rewrite the action Eq.(1) in the vacuum as follows

$$S = \int dt \int r^2 \sqrt{X(r)} dr \, [\tfrac{1}{2}\left(-\tfrac{B''(r)}{X(r)} + \tfrac{B'(r)X'(r)}{2X^2(r)} - \tfrac{4B'(r)}{rX(r)} + \tfrac{2B(r)X'(r)}{rX^2(r)} + \tfrac{2}{r^2} - \right. \qquad (8)$$
$$\left. \tfrac{2B(r)}{r^2X(r)}\right) + F(G) - \alpha \left[G + \tfrac{2}{r^2X(r)^3}\left(-2X(r)B(r)B''(r) - X(r)X'(r)B'(r) + \right.\right.$$
$$\left.\left. 3X'(r)B(r)B'(r) + 2X^2(r)B''(r) - 2X(r)B'^2(r)\right)\right]]$$

The variation with respect to $G$, gives

$$\alpha = \frac{dF(G)}{dG} = F_{,G} \qquad (9)$$

Substituting $\alpha$ into Eq.(8), and by making an integration by part, the Lagrangian takes the following form

$$L(B, B', X, X', G, G') = \sqrt{X(r)} \left[2\left(\tfrac{rB'(r)}{X(r)} + \tfrac{B(r)}{X(r)} - X(r) - \tfrac{rB'(r)X'(r)}{X^2(r)}\right) + \right.$$
$$\left. r^2(F - GF_{,G}) - \tfrac{4}{X^3(r)} G'^{F,GG}(B(r) - X(r))\right].(10)$$

.

Varying with respect to $X(r)$, one obtains the first equation of motion

$$\left(F - GF_{,G}\right) - \frac{4B'(r)G'F_{,GG}}{r^2 X(r)} \left(1 - \frac{3B(r)}{X(r)}\right) \tag{11}$$

$$-2\left(\frac{B'(r)}{rX(r)} + \frac{B(r)}{r^2 X(r)} - \frac{X(r)}{r^2}\right) = 0$$

The variation with respect to $B(r)$ leads to the second equation of motion

$$\frac{2G'^{F_{,GG}} X'(r)}{X^3(r)} \left(X(r) - 3B(r)\right) + 4\frac{B(r)X(r)}{X(r)^2} \left(G'' F_{,GG} + G'^2 F_{,GGG}\right) = 0. \tag{12}$$

By the variation with respect to $G$, we will recover Eq.(7). For a constant Gauss Bonnet invariant solution ($G = G0$), from Eqs.(7) and (11) and fixing $X(r) = 1$ we finally obtain the de Sitter (anti de Sitter) solution

$$B(r) = 1 - \frac{A}{3} r^2, \tag{13}$$

with

$$A = \frac{1}{2}\left(G_0 - \frac{F(G_0)}{F_{,G}(G_0)}\right), \tag{14}$$

and $R_0 = 4\Lambda$ and $G_0 = \frac{8}{3} A^2$.

To have a unique solution for the metric Eq.(5) we must fix one of the metric elements. Let us assume $X(r)$ is a constant.

## 4  Solution with $X(r) = constant$

In the pervious section, we obtain equations of motion of the action Eq.(1) with the static spherically symmetric metric Eq.(5). Now we want to investigate the vacuum solution for the case that, $X(r)$ is a constant. So according to assume $X(r) = constant$, from Eq.(12) one gets

$$F_{,GGG} G'^2 + F_{,GGG} G'' = 0 = \frac{d^2}{dr^2} F_{,G}. \tag{15}$$

then, the following equation will be obtained

$$F_{,G} = ar + b, \tag{16}$$

where $a$ and $b$ are integration constants.

As a simple case, let us choose $X(r) = 1$, so we can rewrite Eqs.(5), (6) and (7) as

$$ds^2 = -B(r)dt^2 + \frac{1}{B(r)}dr^2 + r^2 d\theta^2 + r^2 \sin\theta^2 d\phi^2 \tag{17}$$

$$R = -B''(r) - \frac{-4B'(r)}{r} - \frac{2B(r)}{r^2} + \frac{2}{r^2}, \tag{18}$$

and

$$G = \frac{4}{r^2}[B(r)B''(r) + B'^2(r) - B''(r)]. \tag{19}$$

By considering Eq.(16) and differentiating with respect to $r$ of Eq.(11) and considering $X(r)$ as a constant, will lead to

$$-rB''(r) + \frac{2B(r)}{r} - \frac{2}{r} + 4a\left(B(r)B''(r) + B'^2(r) - 3\frac{B(r)B'(r)}{r} + \frac{B'(r)}{r}\right) = 0 \tag{20}$$

where $\frac{d}{dr}F_{,G} = G'F_{,GG=a}$. The general solution of Eq.(20) is

$$B(r) = -1 + \frac{1}{4a}r. \tag{21}$$

The Ricci scalar and the Gauss-Bonnet terms for this metric are

$$R = \frac{4}{r^2} - \frac{3}{2ar}, \qquad G = \frac{1}{4a^2 r^2} \tag{22}$$

By considering the above $G$, and using Eq.(16), we finally obtain $F(G)$ as following

$$F(G) = (G)^{\frac{1}{2}} + F_0, \tag{23}$$

where $F_0$ is a constant, so we have $R + (G)^{\frac{1}{2}} + F_0$.

The solutions were obtained in Eqs.(13) and (21) are black holes and we have event horizons. There exists a real positive solution of $B(r^+) = 0$ and we obtain

$$r_1^+ = \pm\sqrt{\frac{3}{A}}, \qquad r_2^+ = 4a \tag{24}$$

here $r_1^+$ and $r_2^+$ are the horizon radius of our solutions of Eqs.(13) and (21), respectively.

Hawking temperature of the black hole obtained as following [17]

$$T_H = \frac{1}{4\pi}\left(\frac{dB(r)}{dr}\right)_{r=r_+}. \tag{25}$$

For our solutions and the horizon radius Eq.(24) the temperature is positive, when $\Lambda > 0$ or $G_0 F_{,G}(G_0) > F(G_0)$ for the solution Eq.(13) and $a > 0$ for the other solution.

Now we turn to apply these results to investigate the entropy. The entropy of the dynamical horizon, can be determined by the Noether charge method [18], and one has [19]

$$S_H = \frac{A_H}{4}\left(1 + \frac{4}{r^2}F_{,G}\right). \tag{26}$$

For our solution Eq.(21) one gets

$$S_H = \frac{A_H}{2} > 0. \tag{27}$$

## 5 Conclusion

The static spherically symmetric solutions for $R + F(G)$ gravity models are studied. We introduced the $G_{\mu\nu}^{F(G)}$ as an $F(G)$ gravity tensor and we saw, if $F(G) \propto G$ then $G_{\mu\nu}^{F(G)} = 0$ and the Einstein equations will be recovered. By considering a static spherically symmetric metric and by introducing the Lagrangian multipliers $\alpha$ and using scalar curvature, we obtained the Lagrangian and equations of motion and for the constant Gauss-Bonnet invariant $G = G_0$ the metric led to de sitter solutions with $R_0 = 4\Lambda$ and $G_0 = \frac{8}{3}A^2$. However, for a large class of these metrics we obtained $F_{,G} = \frac{dF(G)}{dG} = ar + b$ which $a$ and $b$ are constant integrations. Substituting $F_{,G}$ into equation of motion, a new metric is obtained and $F(G) = G^{\frac{1}{2}} + F_0$ as a function of $G$. Then the thermodynamics quantities for this metric as a black hole in the framework of the modified Gauss-Bonnet theory of gravity are investigated.

## References

[1] S. Perlmutter et al. [Supernova Cosmology Project Collaboration], Astrophys. J. 517, 565 (1999).

[2] C. L. Bennett et al., Astrophys. J. Suppl. 148, 1 (2003).

[3] M. Tegmark et al. [SDSS Collaboration], Phys. Rev. D 69, 103501 (2004).

[4] S. W. Allen, et al., Mon. Not. Roy. Astron. Soc. 353, 457 (2004).

[5] P. J. Steinhardt, Critical Problems in Physics (1997), Princeton University Press.

[6] T. Padmanabhan, Phys. Repts. 380, 235 (2003); E. J. Copeland, M. Sami and S. Tsujikawa, Int. J. Mod. Phys. D 15, 1753 (2006); Y -F. Cai, E. N. Saridakis, M. R. Setare, J. Q. Xia, Phys. Rep. 493, 1, 1-60 (2010).

[7] C. Brans and C. H. Dicke, Phys. Rev. 124, 925 (1961).

[8] S. Nojiri and S. D. Odintsov, Int. J. Geom. Meth. Mod. Phys. 4, 115 (2007).

[9] S. Nojiri and S. D. Odintsov, arXiv:0801.4843 [astro-ph]; arXiv:0807.0685 [hep-th]; T. P. Sotiriou and V. Faraoni, Rev. Mod. Phys. 82, 451-497 (2010); F. S. N. Lobo, arXiv:0807.1640 [grqc];
S. Capozziello and M. Francaviglia, Gen. Rel. Grav. 40, 357 (2008);
M. R. Setare, Int. J. Mod. Phys. D17, 2219, (2008); Astrophys. Space Sci.326, 27, (2010).

[10] I. Antoniadis, J. Rizos and K. Tamvakis, Nucl. Phys. B 415 (1994) 497; S. Kawai, M. A. Sakagami and J. Soda, Phy. Lett. B 437 (1998) 284; S. Kawai and J. Soda, Phy. Lett. B 460 (1999) 41; P. Kanti, J. Rizos and
K. Tamvakis, Phys. Rev. D 59 (1999) 083512; N. E. Mavromatos and J. Rizos, Phys. Rev. D 62 (2000) 124004; T. Koivisto and D. F. Mota, Phys. Lett. B 644 ( 2007) 104; T. Koivisto and D. F. Mota, Phys. Rev. D 75 (2007); M. Satoh, S.Kanno and J. Soda, Phys. Rev. D 77 (2008) 023526; M. Satoh and J. Soda, JCAP 0809 (2008) 019.

[11] S. Nojiri, S. D. Odintsov and M. Sasaki, Phys. Rev. D71 (2005) 123509.

[12] S. Nojiri, and S. D. Odintsov, Int. J. Geom. Meth. Mod. Phys. 4 (2007) 115; G. Cognola, E. Elizalde, S. Nojiri, S. D. Odintsov and S. Zerbini, Phys. Rev. D 73 (2006) 084007; S. Nojiri, and S. D. Odintsov, Phys. Lett. B 631 (2005) 1; S. Nojiri, S. D. Odintsov and M. Sami, Phys. Rev. D 74, 046004 (2006); S. Nojiri and S. D. Odintsov, J. Phys. Conf. Ser. 66, 012005 (2007); S. Capozziello, E. Elizalde, S. Nojiri and S. D. Odintsov, Phys. Lett. B 671, 193 (2009);C. G. Bohmer and F. N. Lobo, [arXiv:gr-qc/0902.2982v3]; K. Nozari, T. Azizi, M. R. Setare, JCAP 0910:022, (2009); J. Sadeghi, M. R.

Setare, A. Banijamali, Eur.Phys.J.C64, 433, (2009); J. Sadeghi, M. R. Setare, A. Banijamali, Phys. Lett. B679, 302, (2009).

[13] S. M. Carroll, A. De Felice, V. Duvvuri, D. A. Easson, M. Trodden and M. S. Turner, Phys. Rev. D 71, 063513 (2005).

[14] G. Cognola, E. Elizalde, S. Nojiri, S. D. Odintsov and S. Zerbini, Phys. Rev. D 75, 086002 (2007).

[15] L. Sebastiani, S. Zerbini, arXiv:1012.5230v3 [gr-qc].

[16] A. De Felice, S. Tsujikawa, Living Rev. Rel. (2010).

[17] R.G. Cai, L.M. Cao, Y.P. Hu, Class. Quant. Grav, 26, 155018 (2009).

[18] R. M. Wald Phys. Rev. D 48, (1993) 3427; V. Iyer, R. M. Wald, Phys. Rev. D 50, 846 (1994); T. Jacobson. G. Kang, R. C. Myers, Phys. Rev. D 49, 6587 (1994); I. Brevik, S. Nojiri, S. D. Odintsov, L. Vanzo, Phys. Rev. D 70, 043520 (2004); H. Maeda, Phys. Rev. D 81, 124007 (2010).

[19] H. Mohseni Sadjadi, Phys. Scripta 05, 055006 (2011).